\documentclass[preprint,showpacs,preprintnumbers,amsmath,amssymb]{revtex4}
\usepackage{graphicx}
\usepackage{dcolumn}
\usepackage{bm}
\usepackage{epsfig}
\begin{document}
\hfill {\bf October  2004}

\title{Zero-curvature solutions \\
of the one-dimensional Schr\"{o}dinger equation}

\author{M. Belloni} \email{mabelloni@davidson.edu}
\affiliation{%
Physics Department \\
Davidson College \\
Davidson, NC 28035 USA \\
}

\author{M. A. Doncheski} \email{mad10@psu.edu}
\affiliation{%
Department of Physics\\
The Pennsylvania State University \\
Mont Alto, PA 17237 USA \\
}

\author{R. W. Robinett} \email{rick@phys.psu.edu}
\affiliation{%
Department of Physics\\
The Pennsylvania State University\\
University Park, PA 16802 USA \\
}

\date{\today}

\begin{abstract}
We discuss special $k=\sqrt{2m(E-V(x))/\hbar^2}=0$ ({\it i.~e.}
zero-curvature) solutions of the one-dimensional Schr\"{o}dinger
equation in several model systems which have been used as idealized
versions of various quantum well structures.  We consider infinite well
plus Dirac delta function cases (where $E=V(x)=0$) and
piecewise-constant potentials, such as asymmetric infinite wells
(where $E=V(x)=V_0>0$).  We also construct supersymmetric partner
potentials for several of the zero-energy solutions in these cases.
One application of zero-curvature solutions in the infinite well plus
$\delta$-function case is the construction of `designer'
wavefunctions, namely zero-energy wavefunctions of essentially
arbitrary shape, obtained through the proper placement and choice of
strength of the $\delta$-functions.
\end{abstract}

\maketitle

\section{Introduction}
\label{sec:introduction}

Many important quantum-mechanical systems, such as the hydrogen
atom and the rotational or vibrational states of molecules, whose
study contributed to the early success of quantum theory, are
nicely coincident with a few of the handful of simple
quantum-mechanical problems which can be solved easily, namely the
Coulomb potential, angular momentum states, and the harmonic
oscillator. While textbooks often treat other similarly tractable
bound state problems, many of these models have only recently
found experimental realizations. For example, the quantum
mechanical problem of a particle `bouncing' from a flat surface
under the influence of a constant force ({\it e.~g.} gravity) is a
staple in many texts and is easily soluble in terms of Airy
functions, but experimental evidence for the lowest stationary
state of neutrons in Earth's gravitational field has only recently
been presented \cite{neutron_gravity}. The creation of a wide
variety of non-classical motional states ({\it e.~g.} coherent and
squeezed states) of ions in artificially produced harmonic traps
\cite{meekhof} is another example of the experimental realization
of specific `designer' quantum wavefunctions. Asymmetric quantum
wells (modeled by piecewise-constant potentials with one side at a
higher constant potential, $V_0$, than the other) are simple
examples of systems which can exhibit both quasi-classical and
decidedly wave-like behavior (for $E>V_0$, where the  solutions
are oscillatory in both parts of the well) as well as providing an
easy introduction to quantum tunneling (when $E<V_0$, so that the
wavefunctions on each side have opposite curvature). Besides being
of pedagogical interest, experiments exciting coherent charge
oscillations from just such asymmetric multiple quantum well
structures have been reported \cite{charge_oscillations},  making
use of both types of solutions in the produced wave packet.

Other idealized potential systems, such as the finite square well
\cite{finite_1} and infinite square well with additional Dirac
delta function ($\delta$-function) potentials, either with
Dirichlet \cite{vugalter} or periodic boundary conditions
\cite{cyclic_lattice} - \cite{segre} have been also discussed,
often in the context of wave packet revival behavior or
bouncing/trapping. It has been suggested \cite{luz_2} that such
systems can be experimentally realized using microwave analogs
\cite{microwave} or as the limit of highly localized quantum
wells.  In a number of such recent examples then, the emphasis has
shifted from searching for physical systems with correspondingly
tractable mathematical analogs, to producing experimentally
realizable versions of model potentials which can then exhibit
desired characteristics, including generating very specific
quantum wavefunctions.

In this note, we focus on the observation that a number of the
model systems mentioned above, besides exhibiting
negative-curvature (oscillatory) and positive-curvature
(exponential) energy eigensolutions, can also support
zero-curvature (hereafter ZC) eigenstates, {\it i.~e.}
straight-line solutions. Such solutions are seldom, if ever,
discussed in the literature, or considered in more general
analyses of these idealized systems. Such ZC solutions of the 1D
Schr\"{o}dinger equation, for which
$k=\sqrt{2m(E-V(x))/\hbar^2}=0$ over a finite spatial interval,
can be supported in asymmetric quantum wells consisting of
piecewise constant potentials (where $E=V(x)=V_0>0$ in part of the
well) or in otherwise free systems (where $E=V(x)=0$) where
additional $\delta$-function interactions are present to
accommodate the boundary conditions. We will consider the latter
case below in some detail and briefly discuss aspects of the
former case at the end. We will also make note of the fact that
zero-energy solutions of the infinite square well plus
$\delta$-function case can be used to generate interesting
supersymmetric partner potentials.

\section{Infinite well plus $\delta$-function interactions}
\label{sec:isw_delta}

\begin{figure}
\epsfig{file=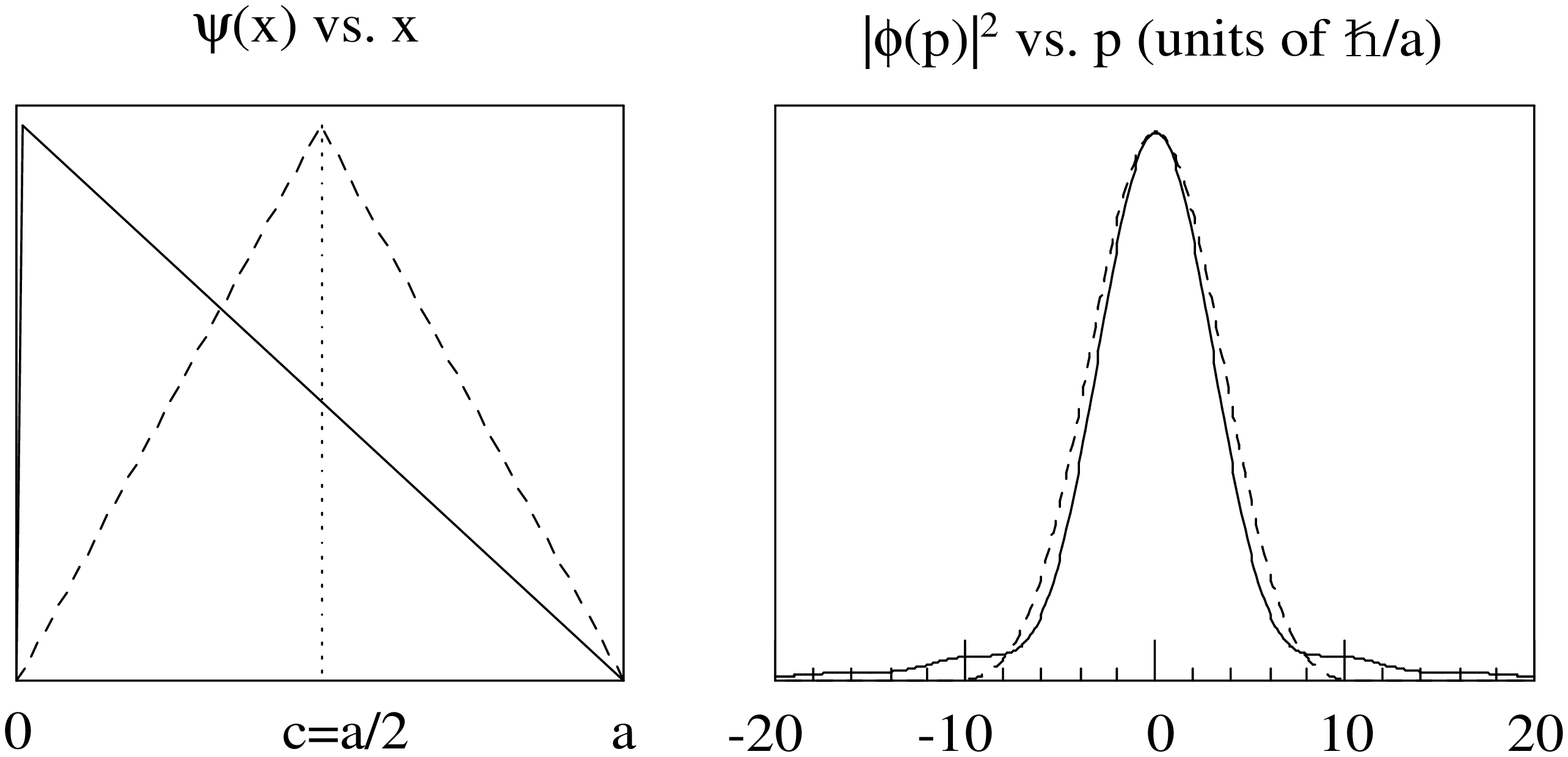,width=7cm,angle=270}
\caption{Plots of $\psi(x)$ versus $x$ for two zero-curvature
wavefunctions in the infinite square well plus $\delta$-function
potential system (left) with the additional $V_{\delta}(x;c)$ potential
located at $c=a/2$ (dashed lines) and $c = 0.01a$ (solid lines). The
corresponding momentum-space probability densities, $|\phi(p)|^2$ versus
$p$ (in units of $\hbar/a$), are shown on the right.
\label{fig:xp}}
\end{figure}

For the standard infinite square well (ISW) defined over the
region $[0,a]$, zero-energy, and hence zero-curvature, solutions
of the Schr\"{o}dinger equation of the form $\psi(x) = A+Bx$ are
formally possible.  However the imposition of the boundary
conditions $\psi(0) = 0 = \psi(a)$ yields the uninteresting result
that $A=B=0$, so that $\psi(x)$ vanishes identically. On the other
hand, a waveform given by
\begin{equation}
      \psi(x) = \left\{ \begin{array}{ll}
                    0 & \mbox{for $x\leq 0$ and $a\leq x$} \\
               Ax/c  & \mbox{for $0 \leq x \leq c$} \\
               A(a-x)/(a-c)  & \mbox{for $c \leq x \leq a$}
                                \end{array}
\right.
\label{first_triangle}
\end{equation}
is a ZC solution of the Schr\"{o}dinger equation which satisfies
the boundary conditions at $x=0$ and $x=a$ (and which is
appropriately normalized if $A = \sqrt{3/a}$), but which has a
discontinuity in $\psi'(x)$, a cusp,  at $x=c$. This form can be
accommodated by the addition of an appropriate strength
(attractive) $\delta$-function potential of the form
$V_{\delta}(x;c) \equiv g \delta(x-c)$ since such a potential
induces a discontinuity  in the derivative given by
\begin{equation}
\psi'(c^{+}) - \psi'(c^{-}) = \left(\frac{2mg}{\hbar^2}\right)\, \psi(c)
\label{cusp}
\, .
\end{equation}
For the waveform in Eqn.~(\ref{first_triangle}),
the appropriate strength, $g_0$, required to yield a ZC solution
\cite{comment} is given by
\begin{equation}
\left(-\frac{A}{(a-c)}\right)
-
\left( \frac{A}{c}\right)
=
\left(\frac{2mg_0}{\hbar^2} \right)\, A
\qquad
\quad
\mbox{or}
\quad
\qquad
 g_0 = - \frac{\hbar^2}{2m}\frac{a}{c(a-c)}
\end{equation}
which is independent of the normalization, $A$, and only depends
on the shape of the triangular waveform.  The interaction
strength required diverges as $c\rightarrow 0$ and $c \rightarrow a$
as expected since the resulting waveform has an increasingly rapid
spatial variation, implying arbitrarily large kinetic energy, as the
waveform is forced to jump almost discontinuously from zero to $A$
(or from $A$ to zero)  near one  of the two boundaries, as seen in
Fig.~\ref{fig:xp}.

This observation can be confirmed more quantitatively by calculating the
expectation values of the potential and kinetic energies in this modified
ISW system. The expectation value of the potential energy is given by
\begin{equation}
\langle V_{\delta}(x;c) \rangle
= \int_{0}^{a} \, V_{\delta}(x;c)\,|\psi(x)|^2\,dx
=
g_{0} |\psi(c)|^2
= - \frac{3\hbar^2}{2mc(a-c)}
\, .
\label{potential_exp}
\end{equation}
The expectation value of kinetic energy can be expressed in one of
two equivalent forms, namely
\begin{equation}
\langle \hat{T} \rangle = \int \psi^{*}(x)\,
\left(\frac{\hat{p}^2}{2m}\right)\, \psi(x)\,dx = -
\frac{\hbar^2}{2m} \int \psi^{*}(x)\, \frac{d^2 \psi(x)}{dx^2}\,dx
\label{kinetic_form_1}
\end{equation}
or
\begin{equation}
\langle \hat{T} \rangle  = \frac{\hbar^2}{2m}
\int\left|\frac{d\psi(x)}{dx}\right|^2\,dx
\label{kinetic_form_2}
\end{equation}
where the expression in Eqn.~(\ref{kinetic_form_2}) is obtained
from Eqn.~(\ref{kinetic_form_1}) by using an integration by parts.
Use of the form in Eqn.~(\ref{kinetic_form_2}) immediately gives
\begin{equation}
\langle \hat{T} \rangle
= \frac{\hbar^2}{2m}\left[ \left(\frac{A}{c}\right)^2 (c-0)
+ \left(-\frac{A}{a-c}\right)^2(a-c)\right]
=+\frac{3\hbar^2}{2mc(a-c)}
\label{kinetic_exp}
\end{equation}
which when combined with  Eqn.~(\ref{potential_exp}) gives
$\langle V_{\delta} \rangle + \langle \hat{T} \rangle = 0$,
consistent with a zero-energy solution. To make use of the form in
Eqn.~(\ref{kinetic_form_1}), we note that $d\psi(x)/dx$ can be
written in the form
\begin{equation}
\frac{d\psi(x)}{dx}
= \frac{A}{c} + \left(-\frac{A}{a-c} - \frac{A}{c}\right)\Theta(x-c)
\end{equation}
from which we find
\begin{equation}
\frac{d^2\psi(x)}{dx^2} = - \frac{Aa}{c(a-c)}\, \delta(x-c)
\, ,
\end{equation}
giving the same final result as in Eqn.~(\ref{kinetic_exp}).
Either form can then  be used to evaluate the expectation value of $p^2$
and the corresponding spread in momentum, namely
\begin{equation}
\langle p^2 \rangle = \frac{3\hbar^2}{c(a-c)}
\qquad
\qquad
\mbox{and}
\qquad
\qquad
\Delta p = \sqrt{\frac{3\hbar^2}{c(a-c)}}
\, .
\label{momentum_squared}
\end{equation}
The momentum-space wavefunction for this system is also easily derived
and is given by
\begin{eqnarray}
\phi(p) & = & \frac{1}{\sqrt{2\pi \hbar}}
\int_{0}^{a}\, \psi(x)\, e^{-ipx/\hbar}\, dx \nonumber \\
& = &
\sqrt{\frac{3}{2\pi \hbar a}}
\left[\frac{\hbar^2}{p^2}\right]
\left\{ \frac{1}{c}\left(e^{-ipc/\hbar}-1\right)
+
\frac{1}{(a-c)}\left(e^{-ipc/\hbar} - e^{-ipa/\hbar}\right)
\right\}
\label{momentum_space_waveform}
\, .
\end{eqnarray}
The corresponding probability density is then given by
\begin{eqnarray}
|\phi(p)|^2 & = &
\frac{3\hbar^3}{\pi c(a-c)p^4}
\left[
\frac{1}{c}[1-\cos(pc/\hbar)]
+
\frac{1}{(a-c)}[1-\cos(p(a-c)/\hbar)] \right. \nonumber \\
& &
\qquad
\qquad
\qquad
\qquad
\qquad
\left.
-
\frac{1}{a}[1-\cos(pa/\hbar)]
\right]
\label{momentum_space}
\end{eqnarray}
which, despite the $1/p^4$ factor, is well-behaved at $p=0$,
giving $|\phi(0)|^2 = 3a/8\pi \hbar$, independent of $c$. The
expectation value of $p^2$ can also be evaluated using this
$|\phi(p)|^2$ to confirm  the result in
Eqn.~(\ref{momentum_squared}). The momentum-space distribution,
while giving the same $p=0$ value for all values of $c$, does
develop increasingly broad `wings' as $c\rightarrow 0$ and
$c\rightarrow a$ in order to achieve this result, as shown in
Fig.~\ref{fig:xp}.

\begin{figure}
\epsfig{file=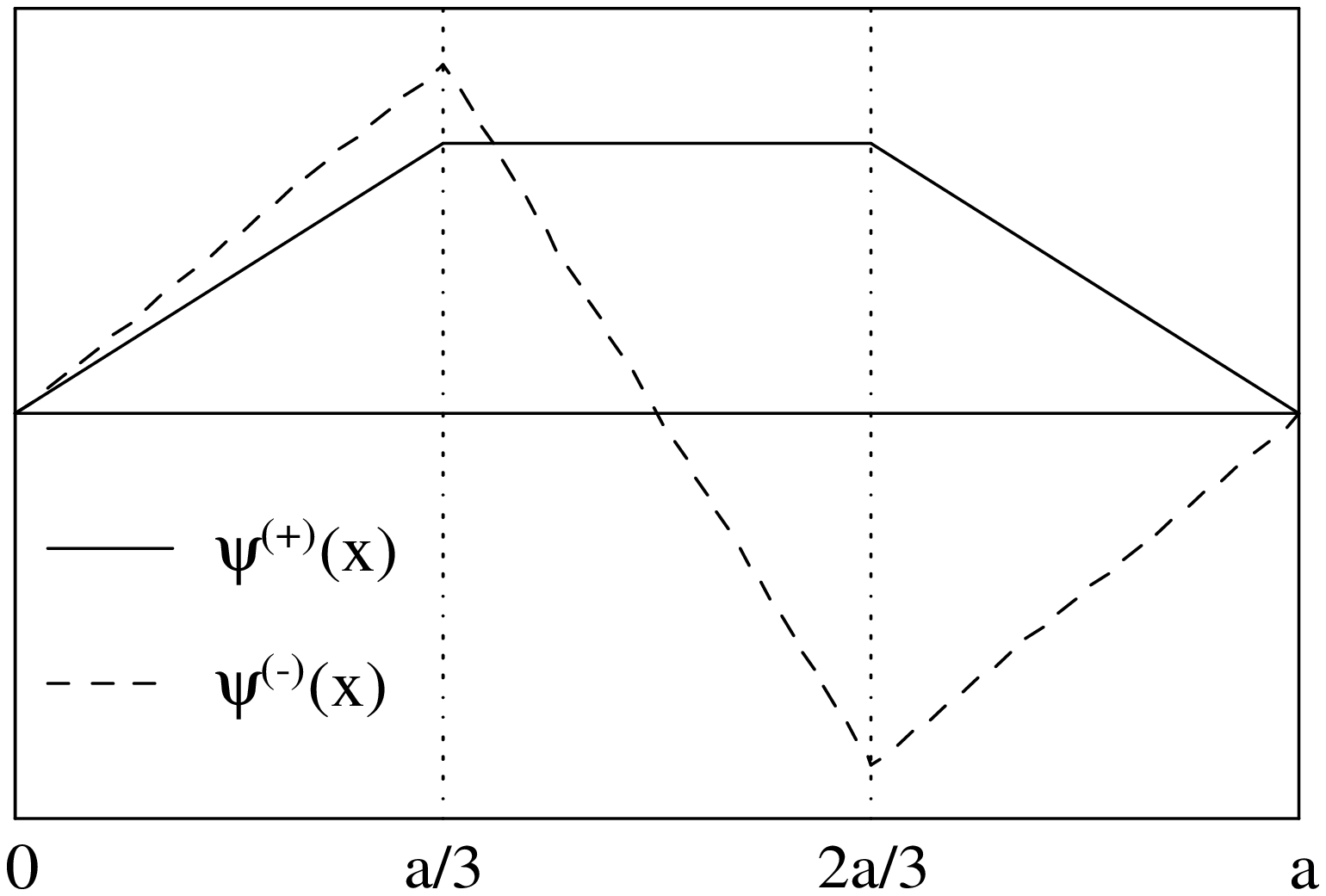,width=7cm,angle=270}
\caption{Plots of $\psi(x)$ versus $x$ for the symmetric (solid)
and antisymmetric (dashed) zero-curvature states in Eqns.~(\ref{symmetric})
and (\ref{antisymmetric}) respectively.
\label{fig:twin}}
\end{figure}

More complex ZC waveforms can then be generated by appropriate
combinations of multiple $\delta$-functions. For example, the symmetric
and antisymmetric (with respect to the center of the ISW) ZC wavefunctions
\begin{equation}
      \psi^{(+)}(x) = \left\{ \begin{array}{ll}
                    0 & \mbox{for $x\leq 0$ and $a\leq x$} \\
               3Ax/a  & \mbox{for $0 \leq x \leq a/3$} \\
               A  & \mbox{for $a/3 \leq x \leq 2a/3$} \\
               3A(1-x/a)  & \mbox{for $2a/3 \leq x \leq a$}
                                \end{array}
\right.
\label{symmetric}
\end{equation}
(where $A = \sqrt{9/5a}$ for normalization) and
\begin{equation}
      \psi^{(-)}(x) = \left\{ \begin{array}{ll}
                    0 & \mbox{for $x\leq 0$ and $a\leq x$} \\
               +3Bx/a  & \mbox{for $0 \leq x \leq a/3$} \\
               +3B(1-2x/a) & \mbox{for $a/3 \leq x \leq 2a/3$} \\
               -3B(1-x/a) & \mbox{for $2a/3 \leq x \leq a$}
                                \end{array}
\right.
\label{antisymmetric}
\end{equation}
(with $B = \sqrt{3/a}$) as shown in Fig.~\ref{fig:twin},
can be obtained using a twin $\delta$-function potential inside the
ISW of the form
\begin{equation}
V_{2}(x) = g
\left[\delta(x-a/3)
+
\delta (x-2a/3)\right]
\label{twin_delta}
\end{equation}
with critical strengths  given by $g_0^{(+)} = -3\hbar^2/2ma$ and
$g_0^{(-)} = -9\hbar^2/2ma$ respectively for these ZC solutions.
These values are consistent with a stronger (more attractive)
potential being required for the antisymmetric case to cancel the
larger kinetic energy, as seen in Fig.~\ref{fig:twin}.
Generalizations to any arbitrary `straight-line' waveform
connecting very general sets of points of the form $(x,\psi(x))$
given by $(0,0)$, $(c_1,A_1)$,...,$(c_N,A_N)$, $(a,0)$ are all
possible with appropriately chosen combinations of additional
$\delta$-function potentials, provided one allows for the
possibility of repulsive ($g>0$) strengths. The kinetic/potential
energy cancellation, yielding $\langle E \rangle =0$, seen in the
single $\delta$-function potential case is easily generalizable to
all such combinations.

An obvious limiting case for comparison is one in which any
desired smooth waveform, $\psi_{0}(x)$, satisfying the ISW
boundary conditions at $x=0$ and $x=a$, can  be approximated by a
large number of short straight-line segments. These segments,
defined by the points $(x_i,\psi_{0}(x_i))$, $i=0,...,N$, where
$N\rightarrow \infty$,  would be matched appropriately using an
increasingly large number of $\delta$-functions. The resulting
critical strengths are given by
\begin{equation}
g_i = \frac{\hbar^2}{2m}
\left(
\frac{\psi_{0}(x_i+\Delta x) - \psi_{0}(x_i)}{\Delta x}
-
\frac{\psi_{0}(x_i) - \psi_{0}(x_i-\Delta x)}{\Delta x}
\right)/\psi_{0}(x_i)
\approx
\frac{\hbar^2}{2m} \frac{\psi''_{0}(x_i)}{\psi_{0}(x_i)}
\end{equation}
so that the effective potential is given by
\begin{equation}
V_{\delta}(x) = \sum_{i}\,g_i\,\delta(x-x_i)
\approx
\frac{\hbar^2}{2m} \frac{\psi''_{0}(x)}{\psi_{0}(x)}
\end{equation}
which reproduces the potential, $V_{0}(x)$, required to produce
such an arbitrary $\psi_{0}(x)$ waveform with vanishing energy,
namely
\begin{equation}
-\frac{\hbar^2}{2m} \frac{d^2\psi_{0}(x)}{dx^2} + V_0(x) \psi_{0}(x) = 0
\, .
\end{equation}

Infinite square well plus $\delta$-function systems have often been discussed
in the context of wave packet revivals \cite{vugalter}, \cite{luz_2},
\cite{segre} where
the time evolution of localized wave packets depends critically on the
detailed energy spectrum, due to the $\exp(-iE_nt/\hbar)$ time
dependence of the individual eigenstates used in the construction of any
initial wave packet. Systems with identical or nearly identical energy
spectra (isospectral) are then expected to exhibit identical
wave packet evolution
and the simplest examples of such related systems are partner potentials
related by supersymmetry \cite{susy}. It is useful to recall
that in cases where a potential
$V_{(-)}(x)$ is known to have a nodeless, ground state eigenstate given by
$\psi_{0}(x)$, arranged to have zero energy ($E_0^{(-)}=0$), then the
supersymmetric partner potential given by
\begin{equation}
V_{(+)} = +V_{(-)}(x) - \frac{\hbar^2}{m}
\frac{d}{dx}\left(\frac{\psi_{0}'(x)}{\psi_{0}(x)}\right)
=
-V_{(-)}(x) +  \frac{\hbar^2}{m}
\left(\frac{\psi_{0}'(x)}{\psi_{0}(x)}\right)^2
\, .
\end{equation}
The $V_{(\pm)}(x)$ partner potentials can also be written in terms of
the superpotential, defined as
\begin{equation}
W(x) =
-\frac{\hbar}{\sqrt{2m}}\left(\frac{\psi_{0}'(x)}{\psi_{0}(x)}
\right) \,,
\end{equation}
in the form
\begin{equation}
V_{(\pm)}(x) = W^2(x) \pm \frac{\hbar}{\sqrt{2m}}W'(x) \, .
\end{equation}
The $V_{(\pm)}(x)$ potentials can be shown to have the identical
energy spectra, except for the $E_0^{(-)}=0$ state of $V_{(-)}(x)$
which is absent in the spectrum of $V_{(+)}(x)$; the energy
eigenfunctions of $V_{(\pm)}(x)$,  the $\psi^{(\pm)}_{n}(x)$, are
then related to each other by generalized raising and lowering
operators. (We are not considering periodic potentials in which it
is possible \cite{dunne} for both isospectral potentials to allow
zero-energy ground state modes.)

Many of the ZC, and hence zero-energy, solutions discussed above
are then immediate candidates for use in generating quite distinct
partner potentials to the ISW plus $\delta$-function potential
systems considered here and elsewhere, which then have the same
energy spectrum and presumably the same pattern of revivals,
trapping, and bouncing discussed previously \cite{luz_1}. For
example,  the symmetric, ZC   solution of the twin-delta function
potential in Eqn.~(\ref{symmetric}), with the critical value of
$g_{0}^{(+)}= -3\hbar^2/2ma$, is the nodeless, zero-energy ground
state of the potential $V_{2}(x)$ in Eqn.~(\ref{twin_delta}) and
so can be used as the $\psi_{0}(x)$ in the construction of a
supersymmetric partner potential. The corresponding $V_{(+)}(x)$
is given by
\begin{equation}
V_{(+)}(x)  =  +\frac{3\hbar^2}{2ma^2}
\left[\delta(x-a/3) + \delta(x-2a/3)\right]
+
{\cal V}(x)
\end{equation}
where
\begin{equation}
      {\cal V}(x) = \frac{\hbar^2}{2m}
\left\{ \begin{array}{ll}
1/x^2 & \mbox{for $0< x< a/3$} \\
0  & \mbox{for $a/3< x < 2a/3$} \\
1/(a-x)^2 &  \mbox{for $2a/3<x<a$}
\end{array} \right.
\end{equation}
and this combination of two repulsive $\delta$-function terms and the
distinctive ${\cal V}(x)$ form will have the same overall spectrum
as $V_{2}(x)$, except for the missing zero-energy ground state.

Finally, similar ZC solutions can be found for the infinite well modified
with periodic boundary conditions (as considered in Refs.~\cite{luz_1}
and \cite{luz_2}) or in the related case of the quantum mechanical
rotor defined by the Hamiltonian $\hat{H} = \hat{L}^2/2I$ with
$\hat{L} \equiv (\hbar/i)(d/d\theta)$ where the eigensolutions are given by
\begin{equation}
      \Theta_{m}(\theta) = \left\{ \begin{array}{ll}
1/\sqrt{2\pi} & \mbox{for $m=0$} \\
\cos(m\theta)/\sqrt{\pi} &  \mbox{for $m>0$} \\
\sin(m\theta)/\sqrt{\pi} &  \mbox{for $m>0$} \\
\end{array} \right.
\end{equation}
which are singly degenerate for $m=0$ and doubly degenerate for
$m\geq 1$. In these cases, ZC solutions are already present in the
unperturbed system, corresponding in the case of the rotor to the
$m=0$ angular wavefunction given by $\Theta_{0}(\theta) =
1/\sqrt{2\pi}$. The addition of a single $\delta$-function
potential cannot provide a solution which is continuous in the
periodic boundary case, but cases  corresponding to the addition
of two or more $\delta$-functions can give a wide variety of ZC
solutions. An example of one class of such solutions with two
$\delta$-functions is that in Eqn.~(\ref{antisymmetric}), or any
such case in which  two equal strength $\delta$-functions are
placed symmetrically around the center of the standard infinite
well, since this type of solution wraps around appropriately under
periodic boundary conditions.

\section{Other examples}
\label{sec:other_examples}

Piecewise-constant potentials also provide examples of local  ZC
wavefunctions for special choices of the well parameters. For
example, the asymmetric infinite square well, defined by
\begin{equation}
      V(x) = \left\{ \begin{array}{cl}
+\infty & \mbox{for $x<-a$} \\
0 & \mbox{for $-a<x<0$ \; (Region I)} \\
+V_0 &  \mbox{for $0<x<+b$ \; (Region II)} \\
+\infty & \mbox{for $+b<x$ }
\end{array} \right.
\end{equation}
can support energy eigenfunctions corresponding to both $E>V_0$ and
$V_0>E>0$ and the conditions on the energy eigenvalues for each case are
 easily derived to be
\begin{eqnarray}
q\tan(ka) + k\tan(qb) = 0
\qquad
&\mbox{with} &
k \equiv \sqrt{\frac{2mE}{\hbar^2}}
\qquad
\mbox{and}
\qquad
q \equiv \sqrt{\frac{2m(E-V_0)}{\hbar^2}} \nonumber \\
\kappa \tan(ka) + k\tanh(\kappa b) = 0
\qquad
&\mbox{where}&
\qquad
\kappa \equiv \sqrt{\frac{2m(V_0-E)}{\hbar^2}}
\end{eqnarray}
for $E>V_0$ and $V_0>E>0$ solutions respectively. Turning the
problem around, either form can then be used to find the condition
on the well parameters necessary to have a zero curvature solution
with $E=V_0$, namely,  by taking the $q\rightarrow 0$ or $\kappa
\rightarrow 0$ limits where one finds
\begin{equation}
\tan(\chi a) = -\chi b
\qquad
\quad
\mbox{with}
\quad
\qquad
\chi \equiv \sqrt{\frac{2mV_0}{\hbar^2}}
\,.
\label{well_condition}
\end{equation}
In this case, the solutions are given by
\begin{equation}
      \psi(x) = \left\{ \begin{array}{ll}
A\sin[\chi (x+a)] & \mbox{for $-a<x<0$} \\
C(1-x/b) &  \mbox{for $0<x<+b$}
\end{array} \right.
\end{equation}
with ZC solutions in the right side of the well. For large values
of $\chi$, the condition on the well parameters in Eqn.~(\ref{well_condition})
reduces to $\chi \approx
 (2n+1)\pi/2a$ (an integral number of half wavelengths plus
an extra quarter wavelength  in the left side of the well). The
probabilities that the particle is found in the left (I) and right
(II) sides of the well approach
\begin{equation}
P_{I} =
\frac{3a}{(3a+2b)}
\qquad
\qquad
\mbox{and}
\qquad
\qquad
P_{II}
= \frac{2b}{(3a+2b)}
\, .
\end{equation}
Supersymmetric partner potentials for the asymmetric infinite well
are also easy to construct in the case where the ZC solution is the
nodeless ground state of the system.

Many similar examples are possible in such piecewise-constant
potentials, including regions where the ZC wave function is
actually flat (in the case of a symmetric infinite well with an
enclosed symmetric barrier of height $V_0$) or as a threshold
solution of the finite square well, consistent with discussions of
Ref.~\cite{poles} in which the poles of the $S$-matrix for the
finite square well are analyzed for bound, threshold, and
scattering states. The analyses of all such cases is
straightforward and will be discussed in a more pedagogical
context elsewhere \cite{other_paper}.

\vskip 1.0cm

\noindent {\bf Acknowledgments}
We would like to thank Laura P. Gilbert for her helpful feedback
in the preparation of this manuscript.
The work of MB was supported in part by a Research Corporation Cottrell
College Science Award (CC5470).

 \newpage

\clearpage

\newpage
\clearpage

\newpage

\begin{thebibliography}{99}
%
%
\bibitem{neutron_gravity} V. V. Nesvizhevsky {\it et al.}, Nature
{\bf 415}, 297-299 (2002); Phys. Rev. D {\bf 67}, 102002 (2003);
{\it ibid}, {\bf 68}, 108702 (2003).
%
%
\bibitem{meekhof} D. M. Meekhof, C. Monroe, B. E. King, W. M. Itano,
and D. J. Wineland, Phys. Rev. Lett. {\bf 76}, 1796 (1996).
%
%
\bibitem{charge_oscillations} A. Bonvalet, J. Nagle, V. Berger, A. Migus,
J. -L. Martin, and M. Joffre, Phys. Rev. Lett. {\bf 76}, 4392 (1996).
%
%
\bibitem{finite_1} V. Venugopalan and G. S. Agarwal,
Phys. Rev. A, {\bf 59}, 1413 (1999); D. L. Aronstein, and C. R. Stroud, Jr.,
Phys. Rev. A {\bf 62}, 022102 (2000); Am. J. Phys. {\bf 68}, 943 (2000).
%
%
\bibitem{vugalter} G. A. Vugalter, A. K. Das, and V. A. Sorokin,
Phys. Rev. A {\bf 66}, 012104 (2002).
%
%
\bibitem{cyclic_lattice} A. C. de la Torre, H. O. M\'{a}rtin, and
D. Goyeneche, Phys. Rev. E {\bf 68}, 031103 (2003).
%
%
\bibitem{luz_1} A. G. M. Schmidt, B. K. Cheng, and M. G. E. da Luz,
Phys. Rev. A {\bf 66}, 062712 (2002).
%
%
\bibitem{luz_2} A. G. M. Schmidt and M. G. E. da Luz,
Phys. Rev. A {\bf 69}, 052708 (2004).
%
%
\bibitem{segre} C. U. Segre and J. D. Sullivan, Am. J. Phys.
{\bf 44}, 729 (1976). This paper was one of the first to describe wave
packet revivals, as well as to consider the effect of $\delta$-function
barriers on such behavior in the context of the infinite square well.
%
%
\bibitem{microwave} U. Kuhl and H.-J. St\"{o}ckman,
Physics E (Amsterdam) {\bf 9}, 384 (2001); U. Kuhl, F. M.  Izrailev, A. A.
Krokhin, and H.-J. St\"{o}ckman, Appl. Phys. Lett. {\bf 77},
633 (2000); U. Kuhl and H.-J. St\"{o}ckman, Phys. Rev. Lett.
{\bf 80}, 3232 (1998).
%
%
\bibitem{comment} One can, of course, solve for the quantized energy
eigenvalues and normalized wavefunctions for general $E>0$ and $E<0$
solutions with the additional $\delta$-function interaction, and both
yield the results here in the limit that $E \rightarrow 0$.
%
%
\bibitem{poles} D. W. L. Sprung, H. Wu, and J. Martorell,
Am. J. Phys. {\bf 64},  136 (1996).
%
%
\bibitem{susy} For a review of supersymmetric quantum mechanics,
see F. Cooper, A. Khare, and U. Sukhatme, Phys. Rep. {\bf 251}, 267 (1995);
{\it Supersymmetry in quantum mechanics}, World Scientific Press, Singapore
(2002).
%
%
\bibitem{dunne} G. Dunne and J. Feinberg, Phys. Rev. D {\bf 57},
1271 (1998).
%
%
\bibitem{other_paper} L. P. Gilbert, M. Belloni, M. A. Doncheski,
and R. W. Robinett, to be submitted for publication.
%
%
\end{thebibliography}
\end{document}